\documentclass[a4paper,fleqn,usenatbib]{mnras}
\usepackage{newtxtext,newtxmath}
\usepackage[T1]{fontenc}
\usepackage{ae,aecompl}
\usepackage{graphicx}	
\usepackage{subfig}
\usepackage{amsmath}	
\usepackage{nicefrac}
\usepackage{xspace}

\usepackage{color}

\newcommand{\tks}{\ensuremath{t_{\rm KS}}}
\newcommand{\Phibh}{\ensuremath{\Phi_{\rm BH}}}
\newcommand{\simlr}{\texttt{MAD-128}\xspace}
\newcommand{\simhr}{\texttt{MAD-192}\xspace}
\newcommand{\simcr}{\texttt{MAD-192-CR}\xspace}
\newcommand{\simsane}{\texttt{SANE-256}\xspace}
\newcommand{\rg}{\ensuremath{r_{\rm g}}}
\newcommand{\bhac}{\texttt{BHAC}\xspace}

\title[Flux tubes in MAD accretion]{Flares in the Galactic center I: orbiting flux tubes in Magnetically Arrested Black Hole Accretion Disks}
\author[O. Porth et al.]{
  O. Porth$^{1}$\thanks{E-mail: o.porth@uva.nl},
  Y. Mizuno$^{2,3}$,
  Z. Younsi$^{4,3}$,
  C. M. Fromm$^{5,3,6}$
\\
$^{1}$Anton Pannekoek Institute for Astronomy, University of Amsterdam, Science Park 904, 1098 XH, Amsterdam, The Netherlands\\
$^{2}$Tsung-Dao Lee Institute, Shanghai Jiao Tong University, Shanghai, 200240, China \\
$^{3}$Institut f\"{u}r Theoretische Physik, Goethe Universit\"{a}t, Max-von-Laue-Str. 1, 60438, Frankfurt am Main, Germany \\
$^{4}$Mullard Space Science Laboratory, University College London, Holmbury St. Mary, Dorking, Surrey, RH5 6NT, United Kingdom\\
$^{5}$Black Hole Initiative at Harvard University, 20 Garden Street, Cambridge, MA 02138, USA\\
Center for Astrophysics — Harvard \& Smithsonian, 60 Garden Street, Cambridge, MA 02138, USA \\
$^{6}$Max-Planck-Institut f\"{u}r Radioastronomie, Auf dem H\"{u}gel 69, 53121 Bonn, Germany
}

\date{Accepted 18-Jan-2021; Received 11-Dec-2020; in original form 05-Jun-2020}
\pubyear{2021}
\begin{document}
\label{firstpage}
\pagerange{\pageref{firstpage}--\pageref{lastpage}}
\maketitle

\begin{abstract}
  Recent observations of SgrA* by the GRAVITY instrument have astrometrically tracked infrared flares (IR) at distances of $\sim 10$ gravitational radii (\rg).  In this paper, we study a model for the flares based on 3D general relativistic magneto-hydrodynamic (GRMHD) simulations of magnetically arrested accretion disks (MADs) which exhibit violent episodes of flux escape from the black hole magnetosphere.  These events are attractive for flare modeling for several reasons: i) the magnetically dominant regions can resist being disrupted via magneto-rotational turbulence and shear, ii) the orientation of the magnetic field is predominantly vertical as suggested by the GRAVITY data, iii) magnetic reconnection associated with the flux eruptions could yield a self-consistent means of particle heating/acceleration during the flare events.
  In this analysis we track erupted flux bundles and provide distributions of sizes, energies and plasma parameter.  In our simulations, the orbits tend to circularize at a range of radii from $\sim 5-40\, \rg$.  The magnetic energy contained within the flux bundles ranges up to $\sim10^{40}~\rm erg$, enough to power IR and X-ray flares. We find that the motion within the magnetically supported flow is substantially sub-Keplerian, in tension with the inferred period-radius relation of the three GRAVITY flares.  
\end{abstract}


\section{Introduction}

Near infrared (NIR) observations of the Galactic center have provided an exciting number of discoveries: foremost the precise measurement of the Galactic Center black hole mass and distance of $M\simeq4.15\times 10^6$ and $\simeq 8.178\,\rm kpc$ through astrometric monitoring of stellar orbits  \cite{SchoedelOttEtAl2002,GhezDucheneEtAl2003,CollaborationAbuterEtAl2019}. Furthermore, the gravitational redshift and post-Newtonian orbit of S2 star was recently measured   \citep{Gravity-CollaborationAbuter2018,CollaborationAbuterEtAl2020} setting tight bounds on the compactness of the central mass.  

In addition to precision measurements of stellar orbits, NIR monitoring has revealed recurring $\sim \times 10$ flux increase flares which last for around an hour and occur roughly four times per day \citep{GenzelSchoedelEtAl2003}.  The peak intensity is $\sim 10^{35}\rm erg\, s^{-1}$ and the emission is strongly polarized with changing polarization angle during the flare \citep{EckartSchoedelEtAl2006,TrippePaumardEtAl2007,ShahzamanianEckart2015}.
Besides IR flares, the Galactic Center is also prone to simultaneous X-ray flares, albeit only one in four IR flares also has an X-ray counterpart \citep{BaganoffBautz2001,PorquetPredehlEtAl2003,HornsteinMatthewsEtAl2007}.
The IR flares (but not the X-ray flares) exhibit substructure down to $\sim 1$ minute and large structural variations on timescales of $\sim 20$ minutes \citep{Dodds-EdenPorquet2009,DoGhez2009}.  
These observations suggest a synchrotron origin of the IR emission from a compact region of size $\sim 3\, \rg$ within $\sim 30\rg$ from the black hole \cite{BroderickLoeb2005,WitzelMartinez2018}.  

Recently, the GRAVITY collaboration has reported three bright flares and astrometrically tracked their flux centroids with an accuracy of $\sim 2 \rg$ \citep[][in the following: G18]{GravityCollaborationAbuterEtAl2018}. The centroid positions and polarization swings with periods of $40-60$ minutes were found to be compatible with a relativistic Keplerian circular orbit at $9 \rg$ \citep{CollaborationBauboeckEtAl2020}.  A striking feature of these observations is that the polarization signature implies a strong poloidal component of the magnetic field in the emitting region (G18).  

GRMHD simulations of radiatively inefficient accretion are quite successful in reproducing many aspects of the galactic center such as spectra, source sizes, some aspects of variability and polarization signatures, yet no consensus model reproduces all observables at once 
\citep{MoscibrodzkaGammieEtAl2009,DexterAgol2010,DibiDrappeauEtAl2012,MoscibrodzkaFalcke2013,ChanPsaltis2015,ResslerTchekhovskoy2016,GoldMcKinney2017,ChaelRowan2018,AnantuaResslerEtAl2020}.
In the IR, large uncertainties are present due to the necessity of including electron heating and likely non-thermal processes in the radiative models \citep{ChaelNarayan2017,ChaelRowan2018,DavelaarMoscibrodzkaEtAl2018}.

Before exploring this large and uncertain parameter space, we here focus on the dynamics that could initiate IR flares like the ones observed by G18.
In the context of the G18 flares, simulations of magnetically arrested disks (MAD) \citep[][]{IgumenshchevNarayanEtAl2003,TchekhovskoyNarayan2011,McKinneyTchekhovskoy2012} are particularly promising:
The simulations show frequent eruptions of excess magnetic flux from the saturated black hole magnetosphere.  As first described by \cite{Igumenshchev2008}, these flux bundles appear as highly magnetised ``blobs'' with a dominant poloidal magnetic field component in the accretion disk \citep[e.g.][]{AvaraMcKinney2016,MarshallAvaraEtAl2018,2019ApJ...874..168W}.  Once the excess flux is re-accreted, a repeating quasi-periodic cycle of outbursts from the black hole is set up.

The environment in which these eruptions occur is turbulent and complex and the mechanism behind the flux escape events is somewhat uncertain.  Yet the process is reminiscent of a Rayleigh-Taylor-like interchange between funnel- and disk-plasma which is triggered once the accumulated magnetic pressure overcomes the ram pressure of the accretion stream \citep[see e.g. the discussion in][]{MarshallAvaraEtAl2018}.  Magnetic reconnection might be involved to promote accretion through the magnetic barrier \citep{IgumenshchevNarayanEtAl2003} or to change topology of funnel field lines through a Y-point in the equatorial plane.

Large-scale simulations of mass feeding in the Galactic Center through magnetised stellar winds have recently been presented by \cite{ResslerQuataertEtAl2019,ResslerQuataertEtAl2020}.  They demonstrate that for a wide range of initial wind magnetisations, the (extrapolated) horizon scale magnetic field is of order of the MAD limit.  Similar to MAD accretion, the inner magnetic field is dominated by the polodial component and magneto-rotational instability (MRI) is either marginally or fully suppressed.  This serves as additional strong motivation to study MAD dynamics in context of SgrA* flares.  

The paper is organized as follows:
in section \ref{sec:results} we first describe the GRMHD simulations and analyze timing properties of various diagnostics in the simulations.  We then elucidate on the flux eruption mechanism and describe our method of flux tube selection for the following statistical analysis. We conclude in section \ref{sec:discussion}.

\section{Results}\label{sec:results}

\subsection{Overall characteristics of the simulations}

In this paper, we discuss GRMHD simulations obtained with \bhac \citep{PorthOlivares2017,OlivaresPorthEtAl2019} \footnote{\url{https://www.bhac.science}} using modified Kerr-Schild coordinates \citep{McKinneyGammie2004} and 2--3 levels of static mesh refinement.  Unless stated explicitly, we use units where $G=c=1$, which for instance sets the length unit $\rg = \rm M$, where $\rm M$ is the mass of the black hole. The simulations are initialised with a hydrodynamic equilibrium torus following \cite{FishboneMoncrief1976} with inner edge at $r_{\rm in}=20\,\rm M$ and density maximum at $r_{\rm max} = 40\,\rm M$ and we use an ideal equation of state with an adiabatic index of $\hat{\gamma}=4/3$.  We perturb the initial state by adding a purely poloidal magnetic field capable of saturating the black hole flux.  The particular vector potential reads:
\begin{align}
  A_\phi \propto \max\left[ \left( \frac{r_{\rm KS}}{r_{\rm in}} \right)^{3} \sin \theta_{\rm KS}\, \exp{(-r_{\rm KS}/400)}\, \left(\rho - 0.01\right), \, 0\right] \,,
\end{align}
where subscript $\rm KS$ indicates ordinary Kerr-Schild coordinates.  The initial magnetic field is weak and scaled such that the ratio of pressure maxima $\beta_{\rm ini} := p_{\rm gas, max} / p_{\rm mag, max}$ adopts a value of $\beta_{\rm ini}=100$.  
\begin{table}
  \caption{
    Overview of the simulations, giving spin, resolution, mass-weighted average Quality factors, domain size.
    \label{tab:simulations} 
  }
  \begin{center}
    \scriptsize
    \begin{tabular}{ccccc}
 ID & a & $N_r\times N_\theta \times N_\phi$ & $\langle Q_r\rangle_\rho
 \times \langle Q_\theta\rangle_\rho \times \langle Q_\phi\rangle_\rho$ & $r_{\rm out}$\\
\hline \hline
\simlr   & $+0.9375$ & $256\times128\times128$ & $18.9 \times 14.4 \times 29.6$ & $2500~\rm M$\\
\simhr   & $+0.9375$ &  $384\times192\times192$ & $27.2 \times 22.4 \times 43.0$ & $2500~\rm M$\\
\simcr   & $-0.9375$ &  $384\times192\times192$ & $25.0 \times 21.2 \times 31.3$ & $2500~\rm M$\\
\simsane & $+0.9375$ &  $512\times256\times128$ & $10.6 \times 11.4 \times 10.3$ & $2110~\rm M$
    \end{tabular}
  \end{center}
  \label{tab:sims}
\end{table}

An overview of the simulations used in this paper is given in Table \ref{tab:sims}.  
With the fiducial (dimensionless) spin value of $a=0.9375$, we discuss two MAD cases with increasing resolutions and one standard and normal evolution (SANE)
case for comparison. In addition, a counter-rotating case with $a=-0.9375$ is shown to investigate the spin dependence of the results. To check for convergence of the simulations, we
quote the mass-weighted MRI quality factors (Q-factors, see section \ref{sec:fluxtubes}) also in Table
\ref{tab:sims}.  
As indicated by Q-factors above 10, all simulations have sufficient resolution to capture the magneto-rotational
instability 
\citep[e.g.][]{HawleyGuanEtAl2011,HawleyRichers2013,SorathiaReynolds2012}.

\subsection{Time series}\label{sec:timing}

\begin{figure}
\begin{center}
\includegraphics[width=0.45\textwidth]{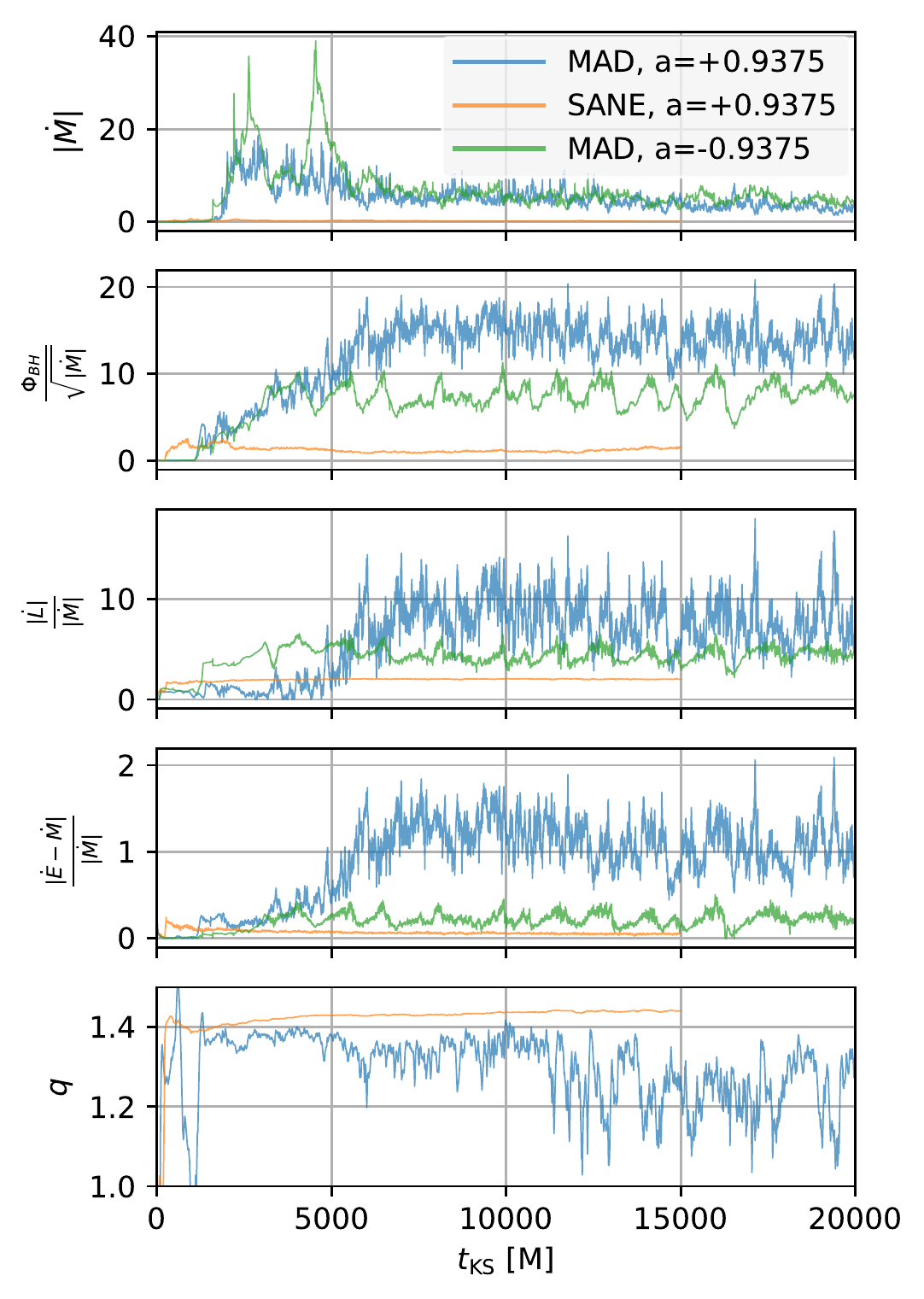}
\caption{Time-series of the MAD accretion runs (blue, green) contrasted with the
  SANE case (orange).  Horizon-penetrating fluxes reach the MAD limits
  at $\tks\simeq 7500~\rm M$. In the co-rotating case, the energy
  extraction is more than $100\%$ of the accretion power at this time.
  Large flux tubes appear at $t\sim12\,000~\rm M$ which coincides with
  strong fluctuations in the rotation index $q$.  Rotation in the SANE
  case on the other hand is only slightly sub-Keplerian with index
  of $q=1.43$ within $\tks\in [5000,10000]~\rm M$.  The rotation index has been omitted for the counter-rotating case at it is not well fit by a power law. 
}
\label{fig:fluxes}
\end{center}
\end{figure}

A time-series of horizon penetrating fluxes following the definitions
of \cite{PorthChatterjeeEtAl2019a} is shown in Figure
\ref{fig:fluxes}.  
A quasi-stationary MAD state is obtained after
$t=7500~\rm M$, where the dimensionless horizon penetrating
magnetic flux $\phi:=\Phi_{\rm BH}/\sqrt{|\dot{M}|}$ reaches the critical
value of $\phi_{\rm max}\approx 15$ for $a=+0.9375$ and $\phi_{\rm max}\approx 8$ for $a=-0.9375$, consistent with \cite{TchekhovskoyMcKinney2012} \footnote{In our system of units, which differs from the commonly employed
definition of 
\cite{TchekhovskoyNarayan2011,TchekhovskoyMcKinney2012,McKinneyTchekhovskoy2012} by a factor of $\sqrt{4\pi}$)}.
Quasi-periodic dips in the horizon penetrating magnetic flux are visible in particular in the counter-rotating case where up to half of the flux is expelled in strong events. The flux is then re-accreted and the expulsion repeats after a timescale of $1000-2000\, M$ which corresponds to $\sim5-10$ hours in the galactic center.  In the co-rotating case, we see weaker flux expulsions and correspondingly the timescale of re-accretion is considerably shorter.  
The normalized accretion power
measured at the event horizon, $|(\dot{E}-\dot{M})/\dot{M}|$, shows an efficiency of up to $\sim 150\%$,
indicating the extraction of spin-energy and is a characteristic property of
the MAD state \citep{TchekhovskoyMcKinney2012}.  
 
Turning to the rotation profiles which are given through the powerlaw:
\begin{align}
  \Omega(r) \propto r^{-q}\,, \label{eq:omega}
\end{align}
with index $q$.
While the SANE case is
compatible with Keplerian motion ($q=1.43$),  once large flux tubes appear around
$t\sim 12\, 000\, \rm M$, the MAD case becomes substantially
sub-Keplerian ($q=1.25$) due to additional magnetic support.  The rotation and
shear will be analyzed further in section \ref{sec:shear}.

It is interesting to consider how accretion of mass and magnetic flux are
inter-related for MAD disks. Naively, if accretion proceeds through
an interchange process, one might expect $\dot{M}(t)$ and $\dot{\Phi}_{\rm BH}(t)$
to be anti-correlated.  This is because dense plasma (increasing $\rm M$) is ``interchanged'' with strongly magnetised funnel
plasma (decreasing $\Phibh$).
However, as pointed out also by
\cite{BeckwithHawley2009}, the black hole mass \textit{must}
increase, whereas magnetic flux can also decrease due to
``escape'' from the black hole and due to the accretion of opposite
polarity field lines and reconnection.
Hence it is not clear whether a correlation between the two properties
should exist at all, as different processes might govern their respective evolutions.

To analyze the accretion process, as a first step, we investigate the auto-correlation of the rates of mass $\dot{M}$, energy- and angular- momentum $\dot{E}$ and $\dot{L}$ as well as the rate of magnetic flux increase
$\dot{\Phi}_{\rm BH}$. The correlation time is defined as the lag when
the autocorrelation assumes a value of $1/e$ and the
time-series is restricted to a time when the simulations are firmly in the MAD state:
$t\in [10\, 000, 15\, 000]\, \rm M$.  This yields a correlation time
for the (detrended) $\dot{M}$ of $t_{\rm corr,\dot{M}} = 47 ~{\rm M} \, {\rm and} \, 65\,
\rm M$ for the fiducial runs \simlr and \simhr, respectively.  These values are consistent
with the decorrelation time of the ray-traced synthetic images used in
the EHT model fitting \citep{CollaborationAkiyamaEtAl2019d}.
Repeating this analysis for $\dot{E}$ and $\dot{L}$ yields similar results.
Quite in contrast to $\dot{M},\dot{E}$ and $\dot{L}$, it turns out that
in our simulations $\dot{\Phi}_{\rm BH}$ is uncorreltated down to the sampling frequency
of $1\rm M$, both for the MAD and SANE cases.  Accordingly, there is no detectable correlation between mass accretion and $\dot{\Phi}$.   This is a strong 
indication that the black hole
flux in the saturated state is subject to a highly intermittent random
process and does not follow the long-term trends seen in the accretion
of mass.

We have further checked for cross-correlations between the aforementioned quantities and note one striking difference between the SANE and MAD cases: in turbulent SANE accretion, the time-series of $\dot{M}$ and $\dot{L}$ are clearly
anti-correlated, meaning low density streams of gas carry higher than average specific angular momentum.
The MAD cases show no clear correlation.  A possible interpretation could be: if angular momentum of low-density flux tubes is removed via large scale stresses in MAD accretion, it is expected that these low density flux tubes carry systematically lower angular momentum when they are accreted.  Hence one would expect a positive correlation between $\dot{M}$ and $\dot{L}$.  If both turbulent MRI accretion and flux tube accretion occur at the same time, likely no correlation is observed.

\begin{figure*}
\begin{center}
\includegraphics[width=0.95\textwidth]{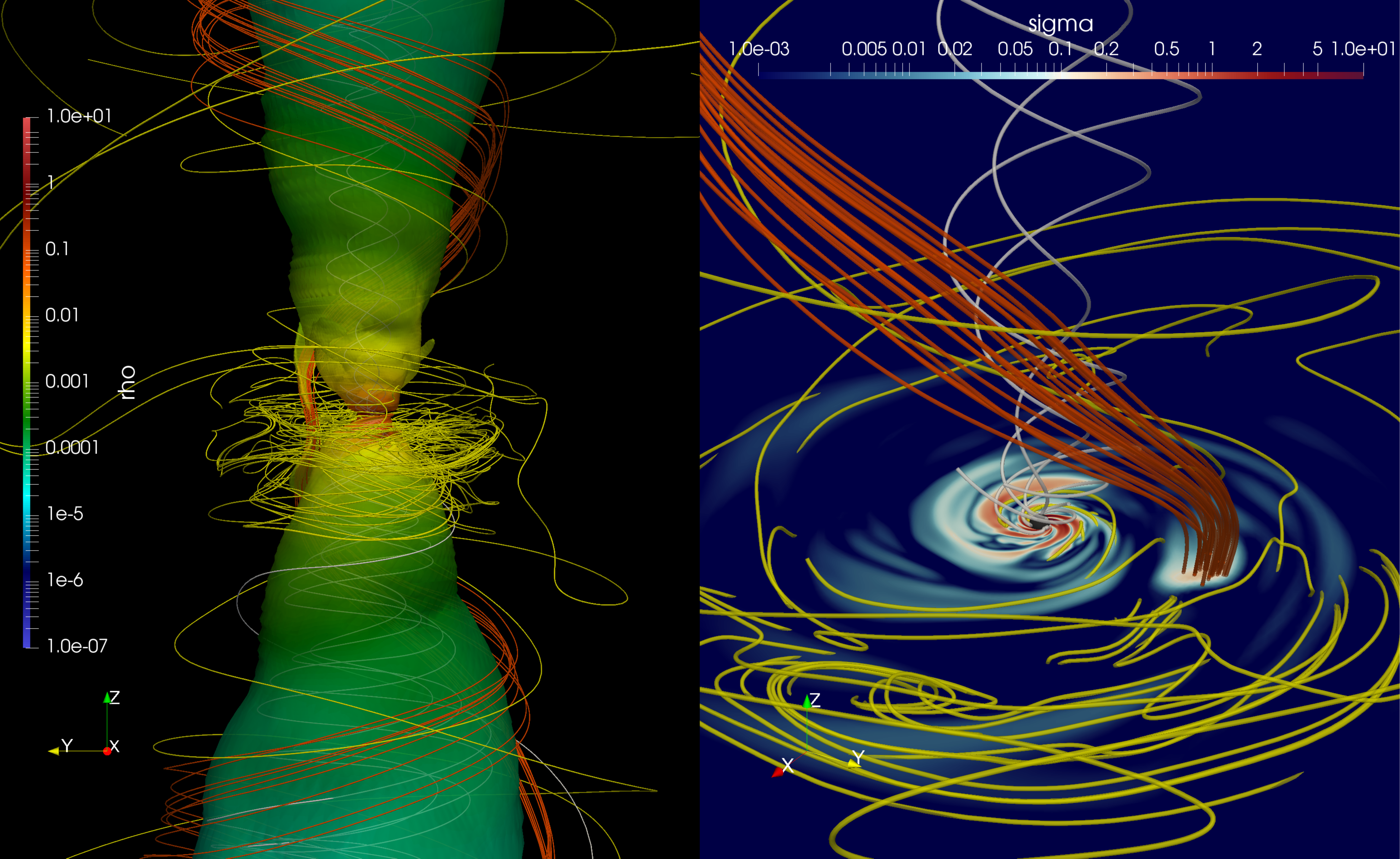}
\caption{Rendering of the different magnetic field components in a MAD
  simulation.  Horizon penetrating field lines (grey), toroidally
  dominated disk fields (yellow) and the expelled flux tube (red).  On
the left panel, we show the iso-contour $\sigma=0.2$ coloured by
density. On the right panel, the mid-plane mangetisation $\sigma$ is
indicated, highlighting also the magnetised flux tube.
}
\label{fig:fieldlines}
\end{center}
\end{figure*}

\subsection{Flux tube selection}\label{sec:fluxtubes}

Flux tubes are dominated by coherent large scale
vertical magnetic fields. They differ substantially from the MRI active regions where the field is sub-dominant and
its geometry is mostly toroidal.
This is visualized in Figure
\ref{fig:fieldlines}, where we chose footpoints rooted on: the black
hole event horizon (white field lines), in the MRI active turbulent disk (yellow lines),
and in a high-magnetization region in the equatorial plane (red). Here we introduce the (hot-) magnetization $\sigma:=b^2/(\rho h)$ which compares the square of the fluid-frame field strength $b^2$ and the enthalpy density of the gas $\rho h$.
Within one scale height of the disk, the flux tube remains nearly
vertical and is subsequently wound up around the jet. Its mid-plane magnetization is $\sigma\simeq 0.5$ and, as the field is strong, the MRI is quenched in the flux tube.  

The suppression of the MRI is quantified by the ``MRI
suppression factor'' which compares the disk scale-height $H$ with the
wavelength of the fastest growing (vertical) MRI mode $\lambda^{(\bar{\theta})}$.  No growth is
expected for wavelengths that do not ``fit'' into the disk diameter,
hence for $S_{\rm MRI}:=\nicefrac{2 H}{\lambda^{(\bar{\theta})}}<1$.
For a quantitative analysis, we define the density weighted averages as:
\begin{align}
  \langle \,\cdot\, \rangle_{\rho}\, (r,\theta,t) := \frac{\int_0^{2\pi} \ (\,\cdot\,)\ \rho(r,\theta,\phi,t) \sqrt{-g} \, {\rm d}\phi}
  {\int_0^{2\pi} \rho(r,\theta,\phi,t) \sqrt{-g}\, {\rm d}\phi}\, , \label{eq:rhoav-rt}
\end{align}
\begin{align}
  \langle \,\cdot\, \rangle_{\rho}\, (r) := \frac{\int_{t_{\rm beg}}^{t_{\rm end}}\int_0^{2\pi}\int_{0}^{\pi} \ (\,\cdot\,)\ \rho(r,\theta,\phi,t) \sqrt{-g} \, {\rm d}\theta\, {\rm d}\phi \, {\rm d}t}
  {\int_{t_{\rm beg}}^{t_{\rm end}}\int_0^{2\pi}\int_{0}^{\pi} \rho(r,\theta,\phi,t) \sqrt{-g}\, {\rm d}\theta\, {\rm d}\phi \, {\rm d}t}\, , \label{eq:rhoav-r}
\end{align}
with $(\,\cdot\,)$ denoting the quantity being averaged, $g$ is the determinate of the four-metric,
and where we set an averaging interval in the quasi stationary state
$t_{\rm beg} = 12000~\rm M$, $t_{\rm end} = 15000~\rm M$.  We measure the (density-) scale height as 
\begin{align}
  \nicefrac{H}{r}\, (r) :=   \langle \,|\pi/2-\theta_{\rm KS}|\,
  \rangle_{\rho}\, (r)\, . \label{eq:scale-height}
\end{align}
The fastest growing mode is evaluated in a co-moving orthonormal
reference frame \citep[][]{Takahashi2008} as: 
\begin{align}
  \lambda^{(\theta)}:= \frac{2\pi}{\Omega \sqrt{\rho h+b^2}} \, b^{(\theta)} \,,
\end{align}
where $\Omega:=u^\phi/u^t$ is the coordinate angular velocity of the fluid.  
We define the average suppression factor as: 
\begin{align}
  \langle S_{\rm MRI} \rangle_{\rho}\, (r,\theta,t) := \frac{
    H(r)\,\langle \Omega \rangle_{\rho}(r)}{\pi \, \langle
    b^{(\theta)}/\sqrt{\rho h + b^2}\rangle_\rho(r,\theta,t)} \,,
\end{align}
and the MRI quality factors as:
\begin{align}
  Q^{(i)} := \frac{\lambda^{(i)}}{\Delta x^{(i)}}, \qquad i\in
  (r,\theta,\phi)\, . 
\end{align}
The mass-weighted averages of the Q-factors within $r<50\,\rm M$ are noted for each run in Table \ref{tab:sims}.  

Effectively, the suppression factor means that MRI does not grow in
magnetically dominated regions.  This can be seen using the
thin disk relation $c_s =\Omega H$ and noting that
$v_A^{(\theta)} =b^{(\theta)}/\sqrt{\rho h + b^2}$ is the
vertical Alfv\'en velocity. Hence
\begin{align}
  S_{\rm MRI} \approx \nicefrac{c_s}{v_A} \approx
  \sqrt{\nicefrac{P_{\rm gas}}{B_z^2}}  \,,
\end{align}
simply compares the sound- and Alfv\'en- velocities or magnetic and
thermal pressure contributions.

To identify flux tubes for further
analysis, we therefore look for regions with dominant vertical field
component and trace the contours where $B_z^2/P_{\rm gas}=1$ in the equatorial
plane.  In addition, to reduce the level of noise in the detection, we
restrict our analysis to flux tubes with a cross-section of at least
$1/(4 \pi r_{\rm h}^2)$ in area, where $r_{\rm h}$ is the radius of the black hole
event horizon.  We have verified, using these criteria, that for the SANE case this comparison does not show any flux tubes.  

\begin{figure*}
\begin{center}
\includegraphics[width=0.95\textwidth]{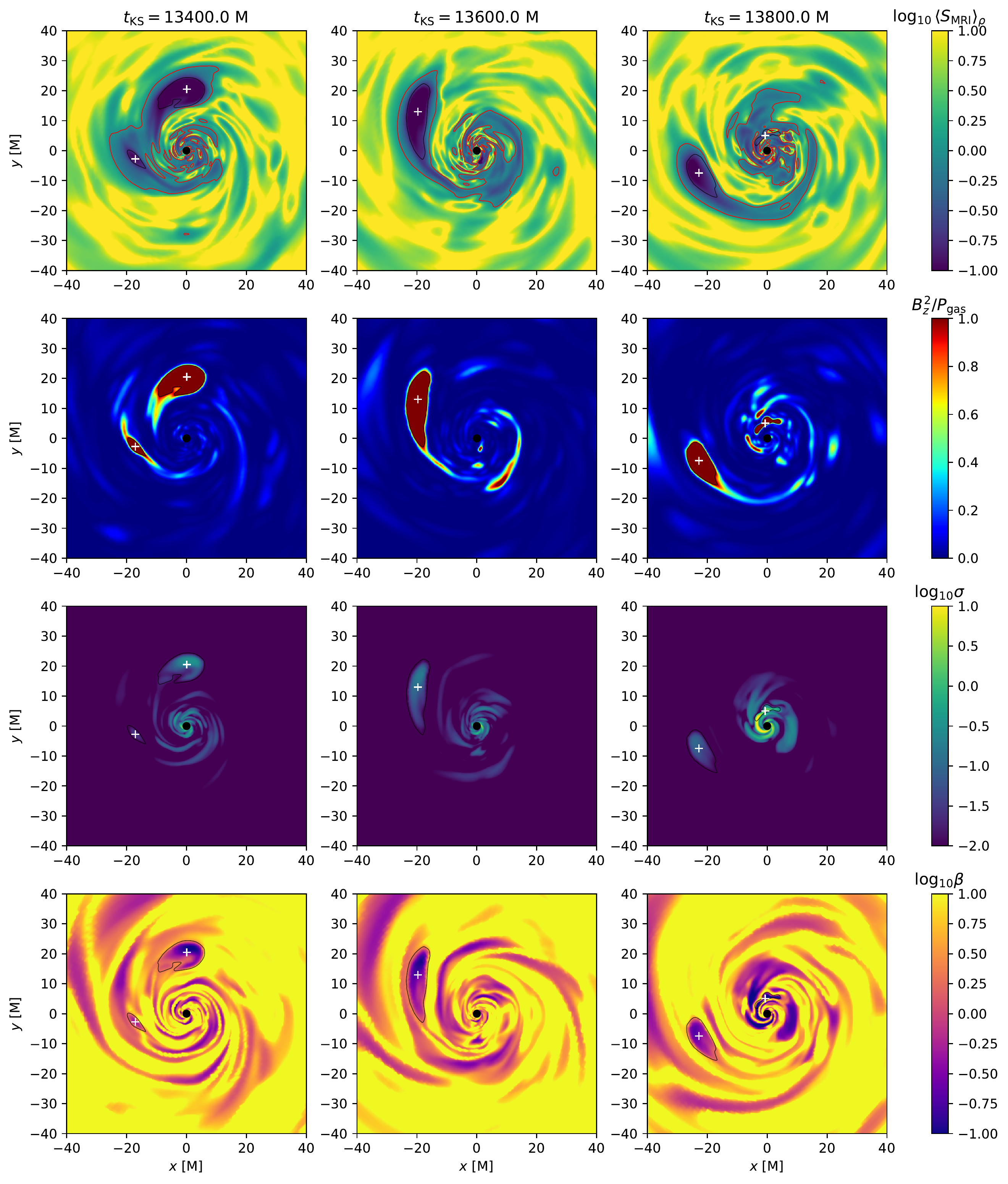}
\caption{Evolution of quantities in the equatorial plane.  We show the density weighted MRI
  suppression factor $\langle S_{\rm MRI}\rangle_\rho$ marking the
  region where MRI is suppressed by the red contour (top panels).
  The second row illustrates $B_z^2/P_{\rm gas}$ used for  extraction
  of flux tubes on the equatorial plane. Magnetisation and
  corresponding plasma-$\beta$ parameter are given in the third and fourth row.  Flux tubes have a dominant vertical field energy $B_z^2$
  compared to the gas pressure and suppress the MRI.  At a radius of $\sim
  20~\rm M$ a large flux tubes performs a circular orbit while shearing
  out in the differentially rotating flow.  
  In the lower three panels, contours mark the detection threshold and crosses the flux centroid
  position. 
}
\label{fig:blobs}
\end{center}
\end{figure*}

Figure \ref{fig:blobs} illustrates the properties of these
regions for three consecutive times for simulation \simlr.  
The flux tubes selected in this fashion have dominant out-of-plane magnetic fields which were checked by tracing their field lines as in Figure~\ref{fig:fieldlines} for several selected cases.
Figure~\ref{fig:blobs} shows that flux tubes coincide with suppressed
MRI (top panels), and have low plasma-$\beta$ and higher than average $\sigma$, as expected (bottom two rows).
It is interesting to note that in both runs \simlr and \simhr, we find that the angle- and time-
averaged $\langle S_{\rm MRI} \rangle_{\rho}\, (r)$ suggests 
MRI suppression within $\sim 10~M$.  However, within this radius, dense
streams of accreting material are frequently found where the MRI can
in principle operate.  

\subsection{Dynamics of flux tubes}\label{sec:dynamics}

Over time, a flux tube will become more elongated as it shears out in
the differentially rotating accretion flow.
Flux- and mass- conservation for constant scale height yields a simple estimate for the pressure contributions in the flux tube:
\begin{align}
  B_z^2 &\propto \Delta r ^{-4} \,, \\
  P_{\rm gas} &\propto \Delta r^{-2\hat{\gamma}} \,,
\end{align}
where $\Delta r$ is a measure of the size of the flux tube (here 
defined as the radius of the circle having the same surface as the
cross-sectional area of the flux tube).
Hence for any causal $\hat{\gamma}<2$, the magnetic pressure decreases faster than the thermal pressure as the flux tube increases in size. As the flux tube moves outwards, the ambient pressure decreases and pressure equilibrium is obtained via expansion of the tube, hence the flux tube expands and looses its magnetic dominance.  
Shear- and Rayleigh-Taylor induced mixing can also increase the size of the flux tube over time.  
Once distributed over a large area, the flux tube cannot remain magnetically dominated and  is dissipated in the accretion flow.

\begin{figure}
\begin{center}
\includegraphics[width=0.32\textwidth]{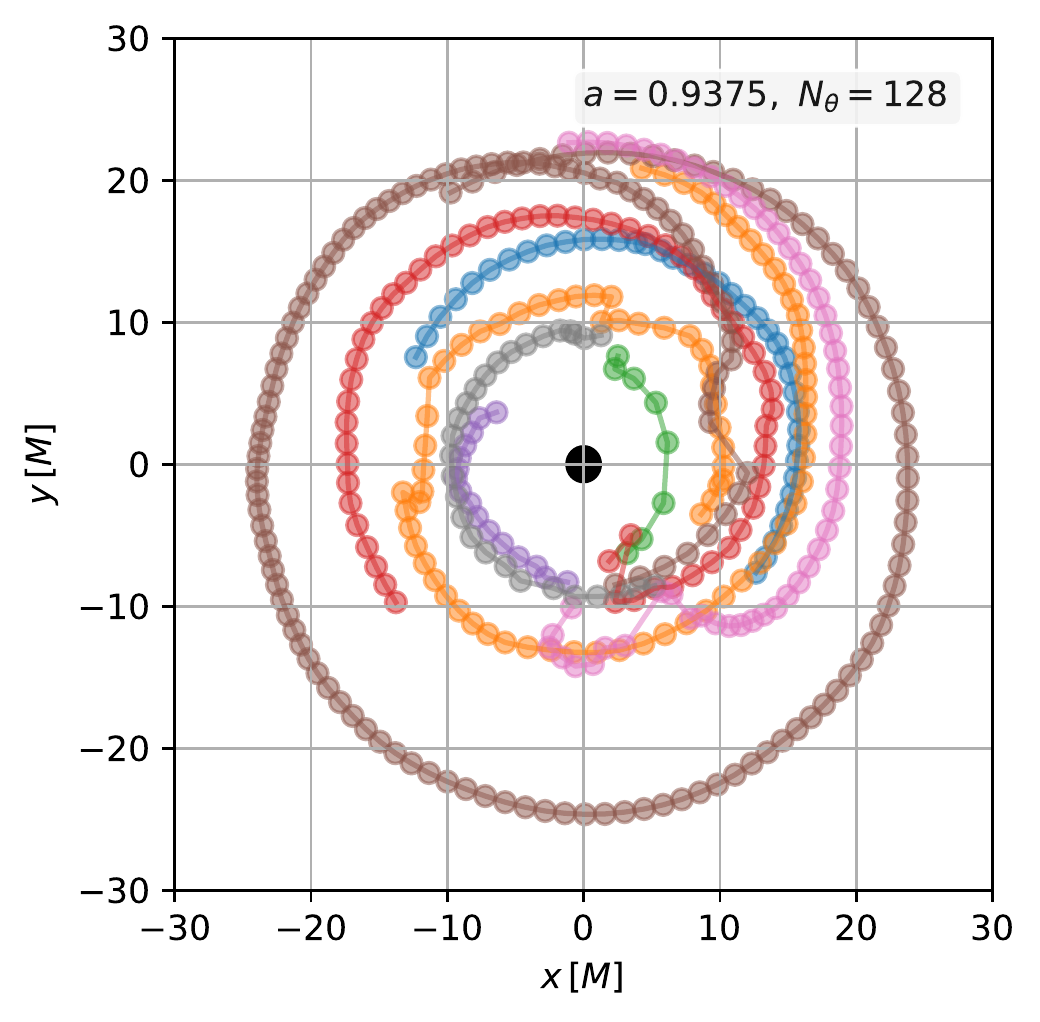}
\includegraphics[width=0.32\textwidth]{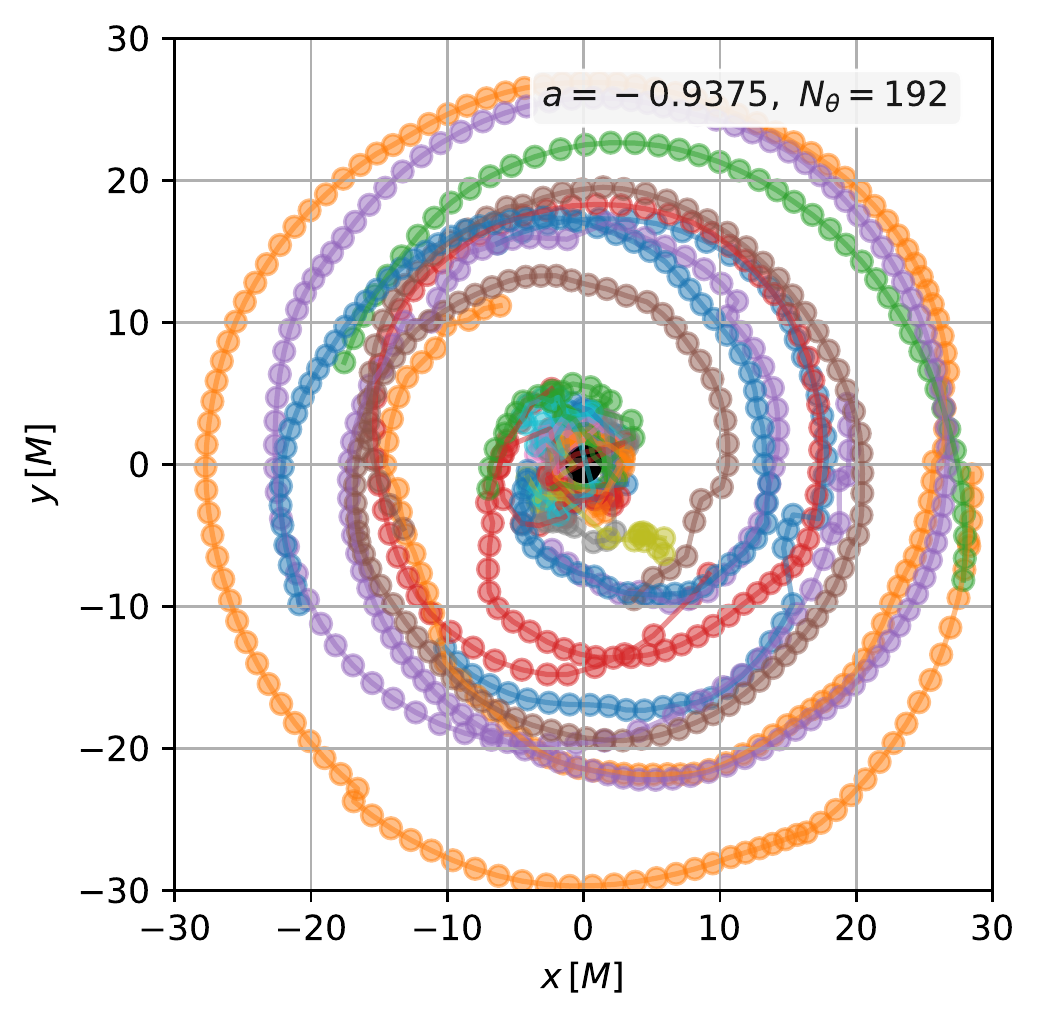}
\caption{Flux centroids positions for the simulations \simlr and \simcr.  We only show data for flux bundles which can be traced for at least one quarter orbit.  The centroid positions spiral outwards and reach nearly circular orbits before the flux bundle dissolves.  We recover a wide range of the circularisation radii and the counter-rotating case also shows tracks in the direct vicinity of the black hole.}
\label{fig:position}
\end{center}
\end{figure}

To analyze the motion of the flux tubes, we compute the centroid of
the magnetic flux in the selected magnetically dominated regions in the equatorial plane, illustrated by ``$+$''
signs in Figure \ref{fig:blobs}.  The centroid motions of robust features which can be tracked for at least one quarter of a circle are illustrated in Figure~\ref{fig:position}. A flux bundle spirals outwards to eventually circularize at what we call its ``circularization radius''. For different flux tubes we recover different circularization radii ranging up to $\sim 40~\rm M$.  Over the considered time interval $t_{\rm KS}\in [12\,000,\,15\,000]~\rm M$, the high-resolution \simhr case shows fewer eruptions than \simlr, yet the parameters of the present features are comparable.  

\begin{figure}
\begin{center}
\includegraphics[width=0.45\textwidth]{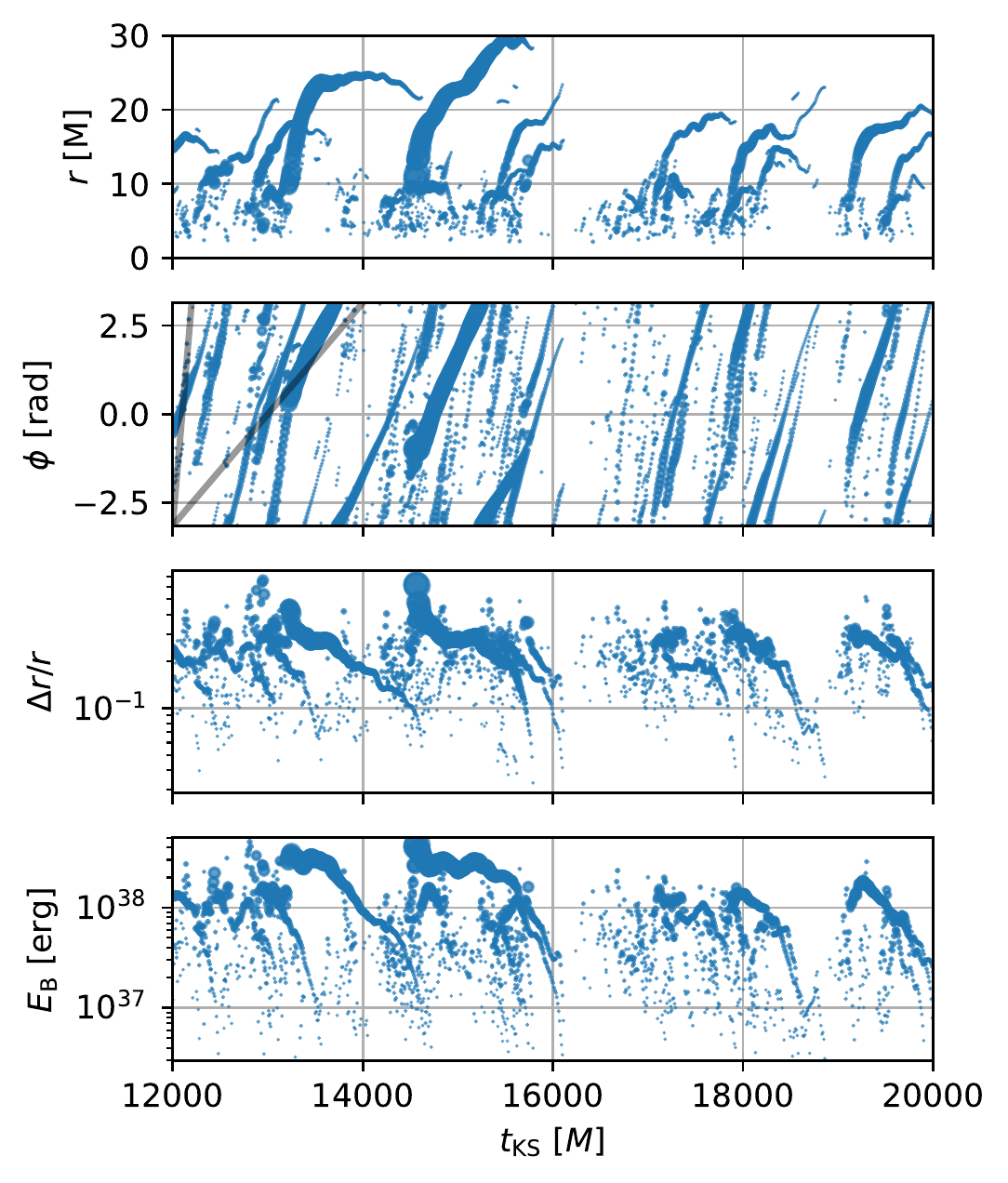}
\caption{Time evolution of the orbiting flux tubes identified in 
  simulation \simlr.  The first two panels represent the coordinates of
  the magnetic flux centroid in the equatorial plane.  In the third
  panel we show the relative size of the flux tube and the fourth panel
  shows the magnetic energy contained in the flux tube within one disk
scale height.  Symbols are scaled according magnetic flux and gray
lines indicating orbital periods of $200~\rm M$ and
$2000~\rm M$ have been added in the third panel to guide the eye.}
\label{fig:orbits1}
\end{center}
\end{figure}

Key parameters of the flux tube evolution are summarized in Figure \ref{fig:orbits1} where we show the coordinate values $r,\phi$, the relative size of the
magnetically dominated region, $\Delta r/r$, and the magnetic energy contained
within one density scale height, $E_{\rm B}$.  To compute the
normalization for the latter, we perform ray-tracing radiative
transfer of the data using the \texttt{BHOSS} code
(\cite{YounsiWu2015}, Younsi et al. 2020) and scale the simulations to
recover the Galactic Center flux of $F_\nu\simeq2.4 ~ \rm Jy$ observed
at an EHT frequency of $230\,\rm GHz$ \citep{DoelemanWeintroub2008}.
We apply an inclination of $45\rm deg$ and ``standard parameters'' from the M87 modeling
\citep{CollaborationAkiyamaEtAl2019d}: $R_{\rm high}=10$ and adopt a
high-sigma cutoff $\sigma_{\rm cut}=1$.

By tracing the radius and azimuth, it is seen that flux tubes
generally move outwards from the black hole due to the magnetic
tension of the highly pinched fields and slow down their radial motion
to orbit at constant radius between $\sim 5$ and $40~\rm M$.
As flux tubes are eroded by the ambient flow and decrease in
  magnetization, just before they dissolve, the detected (radial-)
  centroid motions become erratic.  This is reflected in the first
  panel of Figure \ref{fig:orbits1} as seemingly inward- or outward moving features observed after circularization.  
The large variance in circularization radii indicates that the final
resting place of the flux bundles is not given by the magnetospheric
radius (which on average lies $\simeq 10$).  Rather, we find that when
the field enters a circular orbit it has also adopted a predominantly
vertical orientation and hence no tension force is available to drive
it out further.

Orbital periods between $200~M$ and $2000~M$ are recovered in our
simulations, depending on the radial location of the fields.  
Due to the increase of plasma-$\beta$ (and thus decrease
of the tracer quantity $B_z^2/P_{\rm  gas}$), the inferred size and
magnetic energy gradually decrease until the flux tube is no longer detected as a magnetically-dominated region. At its maximum, the magnetic energy reaches $\sim 5\times 10^{38}~\rm erg$ for both the \simlr and \simhr simulations.

\subsection{Distributions}\label{sec:distributions}
\begin{figure}
\begin{center}
\includegraphics[width=0.45\textwidth]{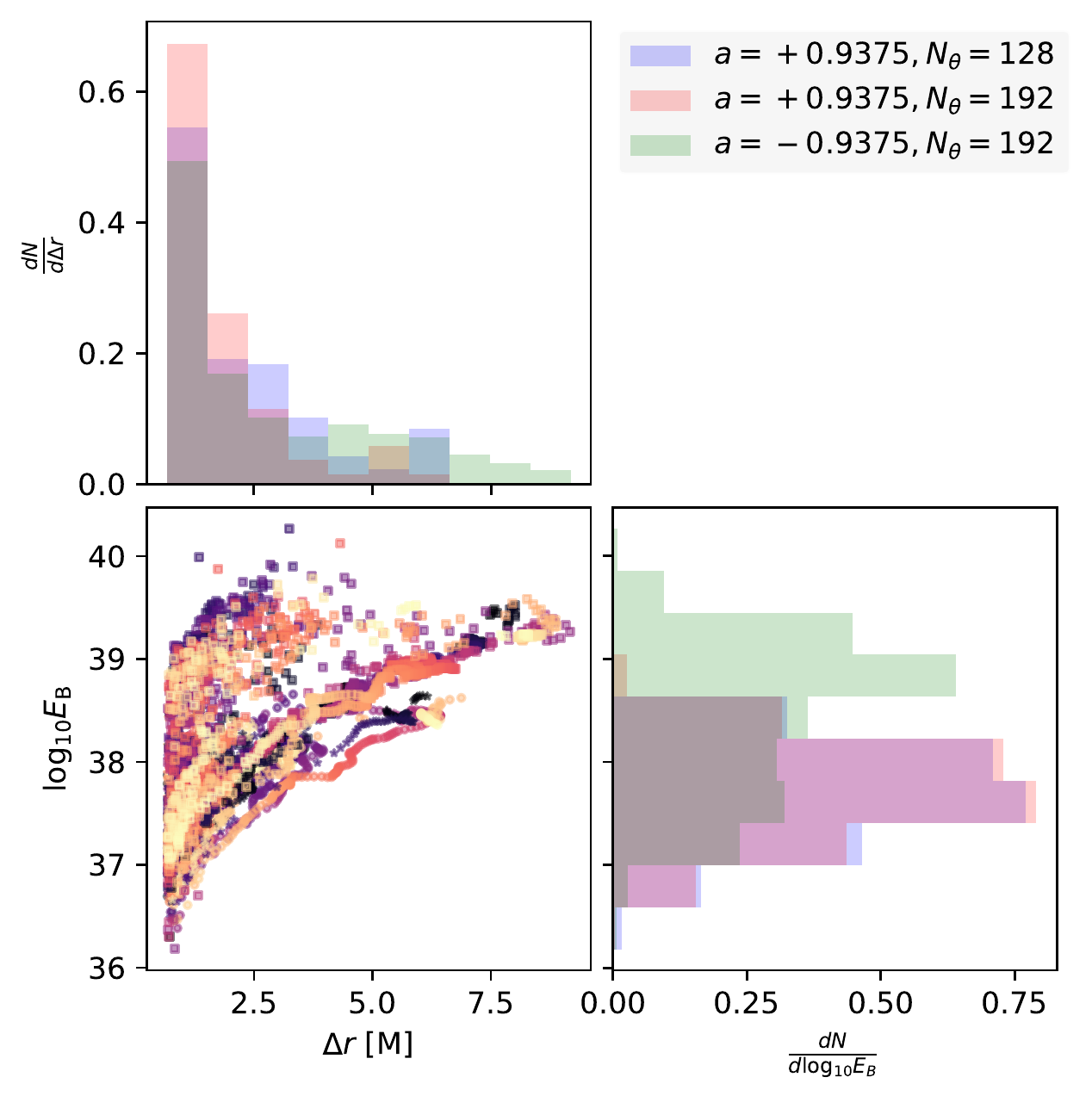}
\includegraphics[width=0.45\textwidth]{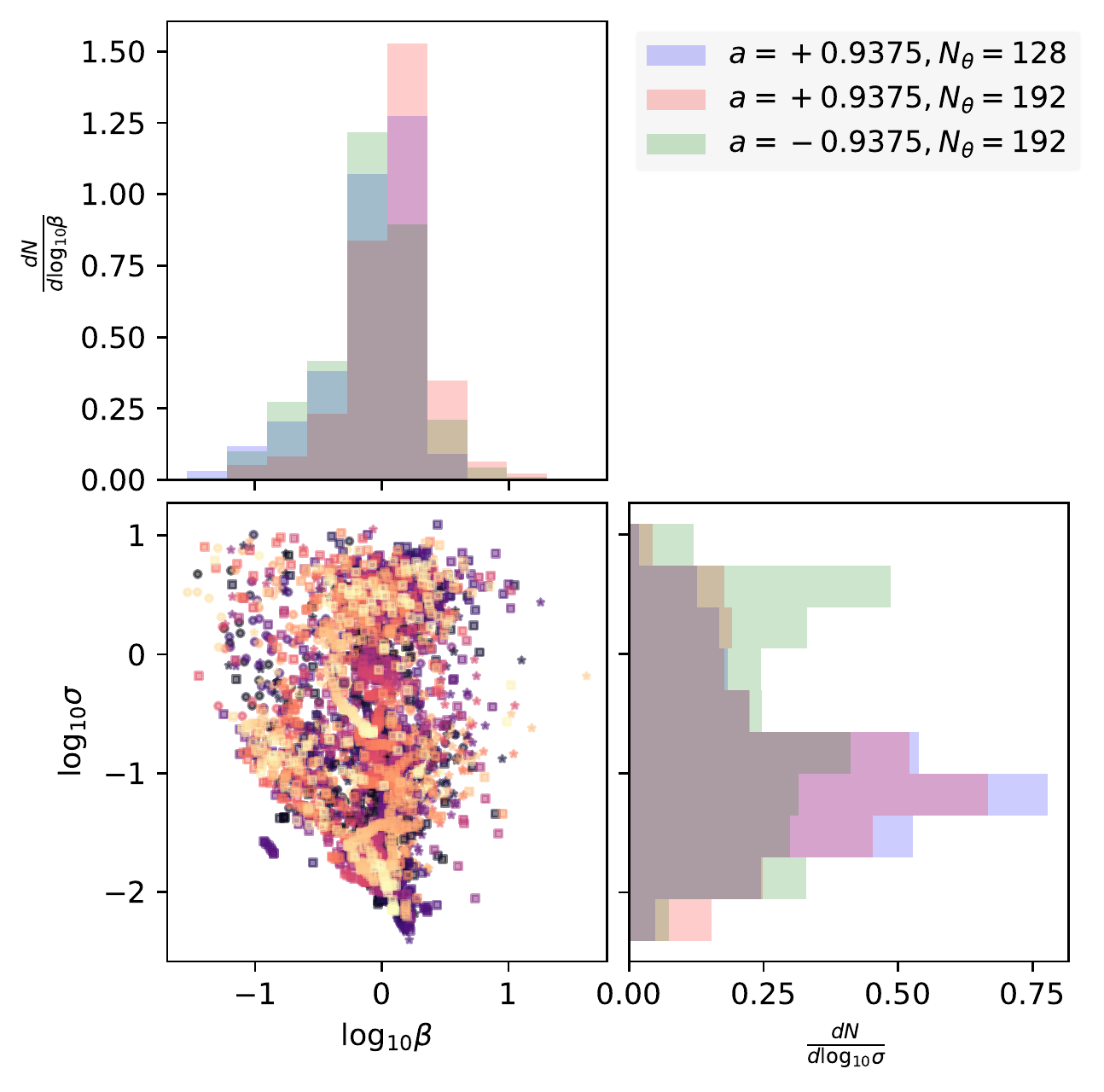}
\caption{Distribution of the flux tubes identified in the MAD
  simulations.  Most of the flux tubes
  have small radii $\lesssim 2 ~\rm M$ and magnetic energies of
  $\sim 10^{38}~\rm erg$ for the co-rotating case and $\sim 10^{39}~\rm erg$ for the counter-rotating case.  Typically, $\beta$ is of order unity and
  magnetisations $\sigma\sim 0.1$ although also highly magnetised cases
$\sigma\in [1,10]$ are observed in particular in the counter-rotating case which show a much broader distribution.}
\label{fig:histograms}
\end{center}
\end{figure}

The instantaneous distributions of various flux
tube properties are shown  in Figure~\ref{fig:histograms}.  Small
flux tubes with $\Delta r\sim 0.5~{\rm M}-1~\rm M$ close to the detection cutoff dominate in number, but sizes of up to $7~\rm M$ are recovered.  As shown in the top panel of Figure~\ref{fig:histograms}, the range of magnetic energies span over two orders of magnitude, from $3\times 10^{36}\rm~erg$ to $3\times 10^{38}\rm~erg$ in the co-rotating case and ranging up to $10^{40}\rm~erg$ in the counter-rotating case.  The most probable magnetic energy of a flux tube is $\simeq5\times 10^{37}\rm erg$ (co-rotating) respectively $\sim 10^{39}\rm ~erg$ (counter-rotating) and large flux tubes are only found with high magnetic energies.
However, smaller flux tubes are found at all magnetic energies.  

Turning to the average plasma-$\beta$ and magnetisation $\sigma$ of the flux tubes, the distribution of plasma-$\beta$ peaks close to the detection threshold $\beta\simeq 2$ but extends down to $\sim0.1$.  The magnetisations start at $0.01$ with the peak at $\sim0.1$ and an extended tail ranging up to $\sigma=10$.  The counter-rotating case has a broader distribution of $\sigma$ with a second peak at $\sigma\simeq3$.  Efficient particle acceleration via magnetic reconnection requires the plasma to be magnetically dominated, hence $\sigma>1$ and $\beta<1$.  In our sample, we find that this is the case for $\sim10\%$ of the identified features.
We have carried out this analysis for both \simlr and \simhr runs, finding that these results are quite insensitive to the choice of resolution and run. 

\subsection{Shearing analysis}\label{sec:shear}
\begin{figure}
\begin{center}
\includegraphics[width=0.45\textwidth]{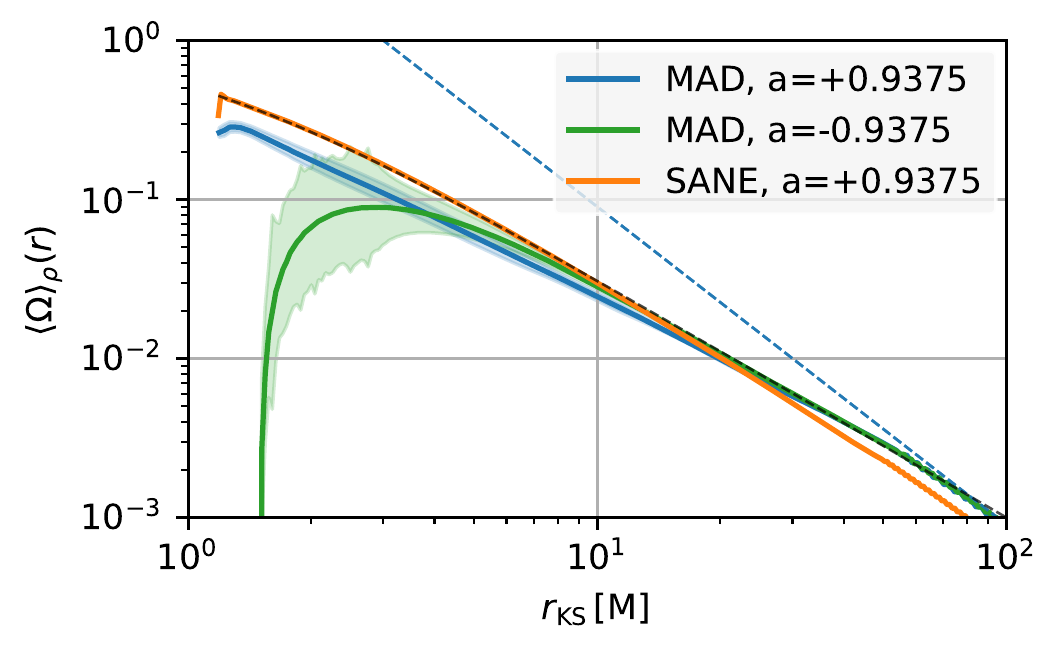}
\caption{Disk- and time- averaged rotation profiles in the
  SANE and MAD runs for an averaging interval of
  $t\in[12000,15000]\,\rm M$  The dashed blue curve indicates the rotation law
  of the initial data $q=2$ and the dashed black curve a relativistic
  Keplerian profile.  The inner regions of the SANE run are consistent
with Keplerian rotation.   Shaded areas underlying the curves
denote the standard deviation of the profiles in time (only visible in
the MAD cases).  }
\label{fig:omega}
\end{center}
\end{figure}

As in any differentially rotating flow, azimuthally advected
features are destined to wind up and loose their coherence.
Naturally, while magnetically dominated and subject to large scale
tension force, flux bundles are not readily sheared however.  In some
sense, they behave more like the spoon that stirs the tea rather than
the milk in it.  Over time however, they will loose magnetic dominace
(see Section \ref{sec:dynamics}) and it is instructive to consider how
long such ``passive tracers'' can remain coherent in a given differential
rotation profile.  This should provide a lower limit to the survival
of the flux tubes against shear.  Given a rotation law of the form
(\ref{eq:omega}), for a feature contained within $[r,r+\Delta r]$, the
inner edge will ``lap'' the slower moving outer component after one
shearing timescale:

  \begin{align}
    t_{\rm shear} := P(r) \left[\frac{1}{ 1 - \left(1+\Delta r/r
        \right)^{-q}} \right] \nonumber \,,
  \end{align}
We show the rotation profile of the simulations $\langle \Omega
\rangle_{\rho}(r)$ in Figure \ref{fig:omega}.  In the SANE case, the
rotation in the inner quasi-stationary regions is described by a
relativistic Keplerian motion $\Omega_{\rm K}:= 1/(a + r^{3/2})$ (dashed black curve) which is fitted by a powerlaw for $r\in[2,20]\,\rm M$ with $q=1.44$.
Due to additional magnetic support, the inner regions of the co-rotating MAD case are sub-Keplerian
with a shallower powerlaw index of $q\simeq 1.25$.  In the counter-rotating case, large departures from Keplerian motion are observed within the ISCO of $\sim 9~\rm M$ and the violent ejection of large flux bundles reflects the large variance of the rotation profile within $\sim 5~\rm M$.  

Varying $q$ in the range $[1.25,1.5]$, however, does not
significantly alter the shearing timescale.
This means that small features with $\Delta r/r<0.1$ can in principle
survive for $\sim 10$ orbital periods, whereas large features with
$\Delta r /r \sim 1$ would be smeared out after roughly one orbit.
For the majority of the detected features with relative sizes $\Delta
r /r \in [0.1,0.3]$, differential rotation allows several orbits
before the features would be fully smeared out due to shear.

\subsection{Orbital periods}\label{sec:periods}

The three astrometrically tracked flares from 2018 reported by \cite{GravityCollaborationAbuterEtAl2018} have shown motion on scales of $\sim 10~\rm M$.
Although only one orbit appears closed, within the measurement errors, all three flares can be explained by a single Keplerian circular orbit with a radius of $9~\rm M$ \citep{CollaborationBauboeckEtAl2020}.  While not statistically significant, \citep{CollaborationBauboeckEtAl2020} note that the sizes of the flux centroid motions appear to be systematically larger than the model predictions.  In other words, the model Keplerian motion at the observed centroid position is too slow compared to the observational data.  Recently, \cite{MatsumotoChanEtAl2020} analyzed the July 22 flare with a broader range of models including marginally bound geodesics and super-Keplerian pattern motion, confirming this finding.  They also find that a super-Keplerian circular orbit with $\Omega = 2.7~\Omega_{\rm K}$ at $r=12.5~\rm M$ yields a better match to the data than the Keplerian orbits.
However, as the measurement errors are substantial, all models are formally acceptable at present.

With the features found in the GRMHD simulations, it is interesting to ask how their orbital periods compare to the data of the flares.
To this end, we need to track features over time in the simulation
data.  Our algorithm works as follows:
\textit{1.} with a cadence of
$10~\rm M$, we obtain the boundary curves of the magnetically
dominated flux tubes in the equatorial plane as described in section
\ref{sec:fluxtubes}.  
\textit{2.} for two consecutive snapshots, we identify a flux tube
from the second snapshot 
with a previously identified flux tube from the first snapshot when their surfaces $S_{1},\,S_{2}$ contained within the
two boundary curves overlap by at least $20\%$.  This
overlap is formally computed as the surface of the intersection between
$S_1,\, S_2$ normalized to their union: $S_1 \cap S_2 / S_1 \cup S_2$.

\begin{figure}
\begin{center}
\includegraphics[width=0.45\textwidth]{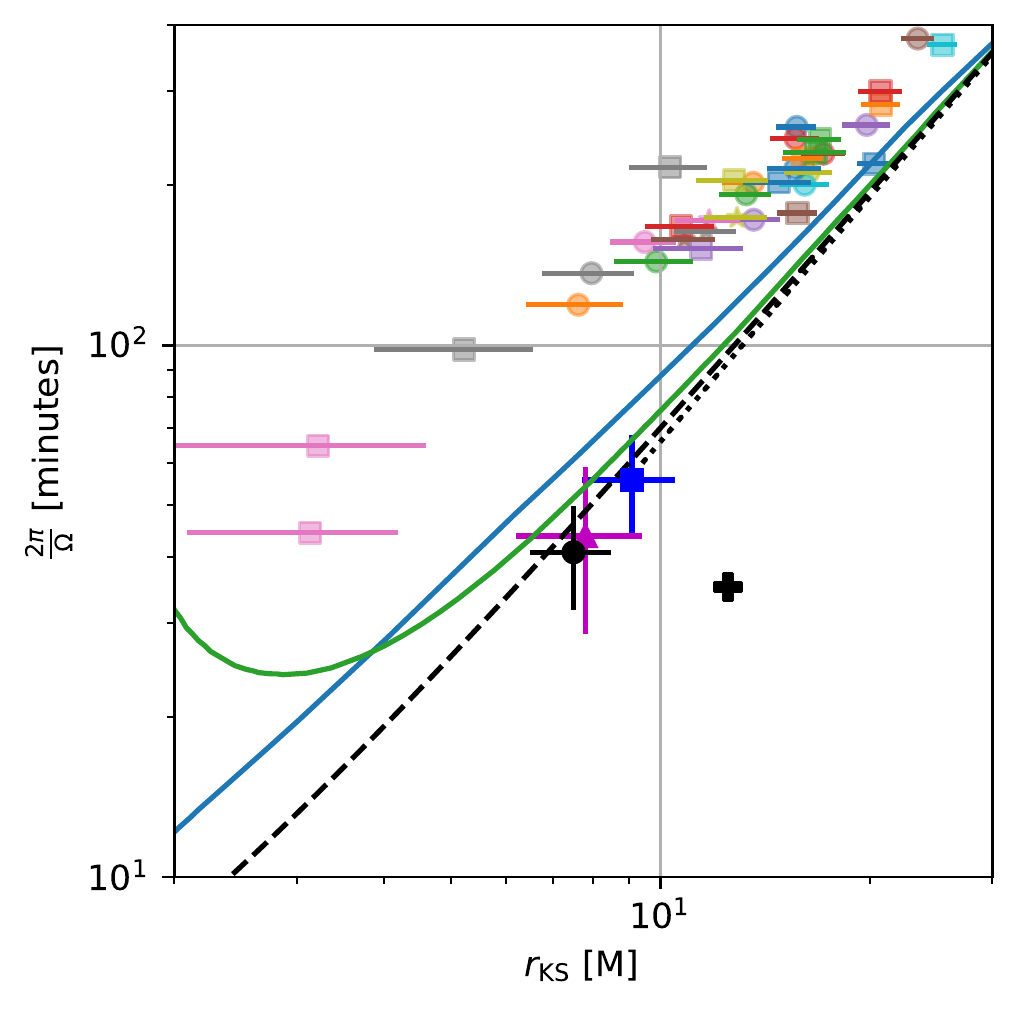}
\caption{Periods against average radius for the features where at least half an orbit can be tracked.  All data is scaled to SgrA*.  Star symbols refer to run \simhr, disks to \simlr and squares to \simcr. 
Blue and green curves show the periods based on $\langle\Omega\rangle_{\rho}(r)$ (cf. Figure \ref{fig:omega}) and dashed (dotted) curve the expected Keplerian profiles for prograde (retrograde) spin.
We also reproduce the datapoints from the analysis of G18 (blue, magenta and black points).
The black + is the super-Keplerian fit to the 2018 July 22 flare from \citet{MatsumotoChanEtAl2020}.  Generally, the orbital periods are sub-Keplerian and even somewhat slower than the density weighted average rotation profile.}
\label{fig:omega-vs-r}
\end{center}
\end{figure}

We verify that this leads to a robust tracking by visually inspecting several test cases.  
Figure~\ref{fig:omega-vs-r} shows the (mean) orbital periods against radius for all features which can be traced for at least $180^\circ$.  To illustrate the radial evolution, we show the standard deviation of the radial coordinate as an error-bar.
As comparison cases, we also overplot: \textit{1.} the datapoints from \cite{GravityCollaborationAbuterEtAl2018} which have been modeled as Keplerian orbits and \textit{2.} the ``super-Keplerian'' pattern motion fit from \cite{MatsumotoChanEtAl2020}.

As shown by the figure, all features are significantly sub-Keplerian and are also slower than the average local rotation velocity by a factor of typically $\sim 2$.  The counter-rotating case shows an abundance of features in the inner region.  These are much slower (up to $\sim 5$ times) than the average flow and exhibit strong radial variation as flux tubes are expelled with large outward velocities.  Hence the fastest feature we could observe has an orbital period of $\sim 40~\rm min$.  Inspecting the instantaneous coordinate velocities of the tracked features, apart from a handful of outliers due to small uncertainties in tracking, here we also do not find any evidence for super-Keplerian motion in the centroid motions.  
Therefore, as MAD flows have generically sub-Keplerian rotation profiles \citep{IgumenshchevNarayanEtAl2003} and flux tubes tend to ``lag behind'' even further due to the magnetic torques exerted by them \citep{SpruitUzdensky2005,Igumenshchev2008}, the MAD model is in tension with the apparent observed fast rotations (respectively large radii of the centroids).

\section{Discussion and Conclusions}\label{sec:discussion}

In MAD disks, low density, high magnetisation flux bundles are frequently expelled from the black hole magnetosphere.  As a candidate scenario for the astrometrically resolved flares observed by \cite{GravityCollaborationAbuterEtAl2018}, we have analyzed the dynamics and energetics of these magnetised regions.

Since the flux bundles coincide with regions of suppressed MRI turbulence, they can remain coherent in the accretion flow for several orbital timescales.  
For the features identified in our simulations, orbital shear
would set an upper limit of $1$ to $10$ orbits, depending on
the size which varies from $\Delta r/r\simeq 0.1-1$.  In practice, a maximum of roughly two
orbits was observed before features ceased to be detected as
  magnetically dominated regions in the accretion flow.  
As strongly magnetized flux tubes are stabilized by the
magnetic tension, orbital shear cannot solely be responsible for the
destruction of flux bundles however. 

Driven outward by magnetic tension from the
radially pinched fields, flux bundles initially move out radially and then
follow circular orbits when the field has straightened out to an
essentially vertical structure.  The model can therefore explain
orbiting features at a range of radii.
We propose that the dissolution of the flux bundle is governed
  by the following two effects: as the outward moving flux bundle seeks pressure
  equilibrium with its surroundings, it is forced to expand which leads to a decrease
  of its magnetic dominance.  Differential rotation and local shear
  induced Kelvin-Helmholtz instabilities \cite[e.g.][]{2019ApJ...874..168W,AntolinYokoyamaEtAl2014}
  are then available to erode the surface of the flux bundle.

By running one counter-rotating case with spin $a=-0.9375$, we have checked that the flux bundles orbiting at relevant radial distances of $\sim 10~\rm M$ do not depend significantly on black hole spin.  Differences arise mainly within $r=2-3~\rm M$, where the counter-rotating case exhibits a steeply declining rotation profile.  However, in the counter-rotating case, the flux bundles are more energetic by an order of magntitude.  This results from two effects: first, the lower radiative efficiency of the counter-rotating case implies that for the same normalizing mm-flux, a higher accretion rate and density is required.  Second, the eruptions found in the counter-rotating case remove a larger fraction of magnetic flux from the black hole (up to $\sim50\%$) resulting in stronger flares.  

We have computed distributions of sizes, magnetisation and energy contained within the flux bundles for the co- and counter-rotating case.  When the simulations are scaled to match the $230\rm GHz$ flux of the Galactic Center, we find that the most probable magnetic energy in the co-rotating case is $\sim 5\times 10^{37}~\rm erg$  and $\sim10^{39}\rm erg$ in the counter-rotating case.  The latter distribution however extends all the way up to $\sim10^{40}\rm erg$.  Given that strong flares radiate up to $10^{38}~\rm erg$ in X-rays  \citep{BaganoffBautz2001,HornsteinMatthewsEtAl2007,BouffardHaggardEtAl2019}, the counter-rotating case has sufficient magnetic energy to allow for a radiative efficiency of a few percent.  
In our sample, in $\sim10\%$ of the cases, we found average plasma parameters with $\sigma>1$ and $\beta<1$, allowing for efficient particle acceleration via magnetic reconnection.

While magnetic reconnection likely plays a role in the expulsion of magnetic flux from the black hole, due to the highly-variable nature of the inner dynamics, it is difficult to identify clear signatures of a topology change of the magnetic field.  Two-dimensional (resistive) GRMHD simulations of MAD disks by \cite{RipperdaBacchiniEtAl2020} on the other hand have shown an episodically forming equatorial current sheet endowed with a plasmoid chain -- a smoking gun of reconnection.  It is an intriguing possibility that flux threading the black hole might escape via reconnecting through this equatorial current sheet as it can provide a means of loading the flux bundles with relativistic particles.  In our simulations, azimuthal interchange instabilities do not allow a strong current sheet to persist and the flow is continuously perturbed by spiral stream of accreting material.
The mechanism that we envision was recently also described in the context of protostellar flares by \cite{TakasaoTomidaEtAl2019}.  In their resistive MHD simulations, reconnection in the equatorial region heats plasma associated with flux removal from the star leading to flare energies consistent with X-ray observations.  
Future resistive 3D GRMHD simulations will be better suited to elucidate the nature of magnetic reconnection in MAD accretion as it enables a parametric exploration of the resistivity.  

An important constraint for the flaring model comes from the period-radius relation of the flares. Whereas the observations by \cite{GravityCollaborationAbuterEtAl2018} suggest Keplerian or even super-Keplerian motion \citep{MatsumotoChanEtAl2020}, since MAD disks are sub-Keplerian and flux bundles tend to lag by an additional factor of $\sim 2$ in the periods (as already pointed out by \cite{Igumenshchev2008}), there is some tension with the current observations.
In this regard, it is important to consider alternative models like the ejected plasmoids studied by \cite{YounsiWu2015,NathanailFrommEtAl2020,BallOezelEtAl2020}.  
In this model, the emission originates from outward moving plasmoids which form due to magnetic reconnection in the coronal regions of the accretion flow. The changed geometry can yield an explanation for the offset between the mean centroid position and the black hole \citep{BallOezelEtAl2020} as well as reconcile the super-Keplerian motion due to finite light-travel time effects \citep{CollaborationBauboeckEtAl2020}.  Whether the model can also explain the energetics, polarization and recurrence time of SgrA* flares remains to be seen.  

It might be worthwile to briefly entertain the possibility of
  a ``confusion scenario'' to explain the apparent super-Keplerian
  nature of the 2018 July 22 flare: as multiple flux tubes can be
  present at the same time, one might wonder what are the chances that
  several independent flares in fact led to the observed
  characteristics.  Our reasoning against it is as follows: if one
  were to observe multiple flaring tubes (not connected by some
  ``pattern motion signal''), it is quite unlikely that 9 out
  of the 10 datapoints presented in G18 for the July 22 flare are
  monotonously increasing their azimuthal angle by a similar amount.
  \footnote{
  Grossly and brazenly simplifying (and dropping the first datapoint
  which breaks the monotonous trend): if 9 flux tubes are present at
  the same time and anyone can light up at any given time, only 9 out
  of the 9!  realizations would yield the observed ordering.  This one in
  40 320 chance does not even include yet the minute likelyhood that 9
  independent large flares are observed within 30 minutes given the
  average flare rate (following poissonian statistics) of around 4 per
  day.}  Of course one can argue whether truly 9 confused flares are
  required to explain the data or whether one could do with less.
  None the less, we belive the odds are strongly stacked against the
  confusion scenario, even more so since as G18 notes, ``all three
  flares can in principle be accounted by the same orbit model''.

It is necessary to discuss several caveats of our analysis.  To detect flux bundles, we look for regions of suppressed MRI and identify features via the ratio of $B_z^2/P_{\rm gas}=1$ in the equatorial plane.  Changing the detection threshold can lead to more or less detected features altering slightly the quantitative distributions measured in Section \ref{sec:distributions}. This has little influence, however, on the inferred motion of the magnetic flux centroid which is used for analysis of orbital periods carried out in section~\ref{sec:periods}.  

In particular for MAD simulations, which show strong magnetisations and steep gradients of plasma parameters within the disk, it is important to check the resolution-dependence of the results \cite{2019ApJ...874..168W}. To this end, we have carried out two simulations of the fiducial case, differing by a factor of $1.5$ in resolution.  We find that the results of our quantifications are generally consistent with each other and have combined both simulations to increase the available statistics of the analysis.  
It is so far unknown what sets the strength of the flux eruptions.  Most likely, thin disks will experience stronger flares \citep{MarshallAvaraEtAl2018}, however a dependence on the initial conditions, e.g. the initial flux distribution in the disk cannot yet be ruled out.  

For a direct comparison with the observational data, one needs to
compute the near-IR intensity and polarization following a ray-tracing
through the simulation data.  This is carried out in a recent parallel
effort by \cite{Dexteretal.2020} who use a long MAD simulation lasting
for $6\times 10^4\rm M$ and apply (thermal) electron heating from (sub-grid) magnetic reconnection models due to \cite{WernerUzdensky2018}.  
As the density in the escaping flux bundles is set by the funnel
floors, emission of the flux bundles themselves is strongly
suppressed.
To estimate the emission from the flux bundles in our study, we have
  carried out the following experiment with the fiducial run:
  we apply a threshold $B_z^2/P_{\rm gas}\ge10$ to only select regions
  of strong vertical flux in the emitting volume.  Applying the
  standard thermal emission model as described in Section
  \ref{sec:dynamics}, in particular $\sigma_{\rm cut}=1$, we
  obtain a contribution of $\sim 0.5\%$ to the $230\rm GHz$ emission
  and the contribution to the $138\rm THz$ flux is even smaller $\sim
  10^{-4}$.
  Raising the high-magnetization threshold to $\sigma_{\rm cut}=25$
  and again normalizing the accretion rate to recover the observed
  $230\rm GHz$ flux increases the total infrared emission by a factor of
  $\sim6$ (from $0.006\pm0.002\, \rm Jy$ to $0.033\pm0.01\, \rm Jy$).
  For comparison: the equivalent increase found by
  \cite{Dexteretal.2020} was a factor of $\sim2$.
  This increase is explained only to a very small part by the added contribution of
  disk-orbiting flux bundles (which have a median $\sigma\simeq 0.1$,
  e.g. Figure \ref{fig:histograms}) but is largely due to the strongly
  magnetized plasma near the jet wall.  In fact, the strong
  dependence of the IR emission on $\sigma_{\rm cut}$ in MAD simulations is a known
  issue which was studied in detail within two-temperature simulations
  by \cite{2019MNRAS.486.2873C} and our results are consistent with
  their findings.  Hence any radiation modeling of IR emission has to
  deal to a smaller or larger degree with the arbitrary truncation via
  $\sigma_{\rm cut}$.

As discussed by \cite{Dexteretal.2020}, when the IR
emission originates from disk plasma (using $\sigma_{\rm cut}=1$), the
emission at the boundary of the flux bundles is enhanced due to
increased heating.  Flux bundles thus stir up the accretion flow and
their motion should also govern the IR centroid on the observational
plane.  The model of \cite{Dexteretal.2020} shares many features with
the observed flares which raises the hope that IR flares might be
explained without invoking high-magnetization material that -- in
simulations -- is plagued by arbitrary floor values and uncertain
electron thermodynamics.  

None the less, the radiative modeling is still complicated by the fact that at least for strong simultaneous X-ray and IR flares, additional physics of non-thermal particle acceleration is required \citep{MarkoffFalcke2001,Dodds-EdenPorquet2009,ChanPsaltis2015a,BallOzel2016}.
Purely thermal models of IR flares relying on gravitational lensing events have also been proposed \citep{DexterFragile2013,ChanPsaltis2015a}, but have difficulty in explaining the required flare amplitude and NIR spectral index.  In fact, spectral modeling indicates that a non-thermal tail in the distribution function is required both in quiescence and during the flare \citep{DavelaarMoscibrodzkaEtAl2018,PetersenGammie2020}.  In particular the flat to inverted spectral index $\nu L_\nu \propto \nu^\alpha$ with $\alpha>0$ during flares \citep{GillessenEisenhauerEtAl2006} is difficult to explain without invoking non-thermal particle acceleration.  The ``redness'' of the spectra produced by thermal distributions was also noted by \cite{Dexteretal.2020} who included reconnection particle heating yet no non-thermal contributions.  
One scenario that comes to mind is that flux bundles can be loaded with relativistic electrons as they violently reconnect in the equatorial region just before a flux escape event.  While this one-off acceleration mechanism might encounter problems explaining X-ray emission from synchrotron electrons which require continuous injection, a telltale signature of such an event would be the outward motion of the flux centroid at the onset of the flare, before it circularises.

We plan to investigate the IR radiative signatures, foremost the flux centroid motion and polarization, incorporating various electron heating and acceleration prescriptions in a follow up publication.

\section*{Acknowledgements}

The authors thank the anonymous referee for raising several interesting points to enhance the discussion of this paper.  
Y.M.~and CMF~are supported by the ERC synergy grant BlackHoleCam: Imaging the Event Horizon of Black Holes (grant number 610058). CMF is supported by the black hole Initiative at Harvard University, which is supported by a grant from the John Templeton Foundation.  Z.Y.~is supported by a Leverhulme Trust Early Career Fellowship. The simulations were performed on GOETHE at the CSC-Frankfurt and Iboga at ITP Frankfurt. This reasearch has made use of NASA's Astrophysics Data System (ADS).

\section*{Data availability}
The data underlying this article will be shared on reasonable request to the corresponding author.

\bibliographystyle{mnras}
\bibliography{astro}

\bsp	
\label{lastpage}
\end{document}